\documentclass{article}

\usepackage[final]{nips_2017}

\usepackage[utf8]{inputenc} 
\usepackage[T1]{fontenc}    
\usepackage{hyperref}       
\usepackage{url}            
\usepackage{booktabs}       
\usepackage{amsfonts}       
\usepackage{nicefrac}       
\usepackage{microtype}      

\newcommand{\Baystate}{Baystate}
\newcommand{\BaystateHealth}{Baystate Health}

\title{Predicting Severe Sepsis Using Text from the Electronic Health Record}

\author{
  Phil Culliton \\
  MultiModel Research\\
  Cambridge, MA  02138 \\
  \texttt{pculliton@mmres.com} \\
       \And
       Michael Levinson \\
       MultiModel Research\\
Cambridge, MA  02138 \\
       \texttt{mlevinson@mmres.com} \\
	         \And
       Alice Ehresman, RN \\
       Baystate Health\\
Springfield, MA  01199 \\
       \texttt{alice.ehresman@baystatehealth.org} \\
       \And
       Joshua Wherry \\
       Baystate Health \\
       Springfield, MA  01199 \\
       \texttt{Joshua.Wherry@baystatehealth.org} \\
       \And
       Jay S.\ Steingrub, MD \\
       Baystate Medical Center \\
       Springfield, MA  01199 \\
       \texttt{jay.steingrub@baystatehealth.org} \\
       \And
       Stephen I.\ Gallant\thanks{Corresponding author}\\
       MultiModel Research \\
       Cambridge, MA  02138 \\
       \texttt{sgallant@mmres.com} \\
}

\begin{document}

\maketitle

\begin{abstract}
  Employing a machine learning approach we predict, up to 24 hours prior, a diagnosis of severe sepsis.  Strongly predictive models are possible that  use only text reports from the Electronic Health Record (EHR), and omit  structured numerical data.  Unstructured text alone gives slightly better performance than structured data alone, and the combination further improves performance.  We also discuss advantages of using unstructured EHR text for modeling, as compared to structured EHR data.
\end{abstract}

\section{Introduction}
Sepsis, a kind of generalized infection leading to organ failure, is a major concern for health providers.  Sepsis causes 20\%-30\% of deaths in hospitals and consumes \$15.4 billion annually in healthcare costs [Henry 2015].  If severe sepsis progresses to septic shock, mortality can run as high as 50\% [Bhattacharjee 2017], with delay in diagnosis and treatment increasing mortality by 7.6\% for every hour of delay.

\subsection{Goals}

Although the clinical definition of severe sepsis involves only \emph{structured data} measurements, the Electronic Health Record also contains potentially useful \emph{unstructured text}:  patient history, progress notes, lab reports, etc.  We would like to use machine learning to automate the understanding of these text notes to make decisions from them, such as ``\emph{Is this patient likely  to satisfy the definition for severe sepsis within the next 24 hours?}''  However, in general, it is difficult for computer software to learn to make decisions from text.  The result is that unstructured notes in the EHR are severely under-utilized for computational purposes [Ohno-Machado 2011].

Working with data from {\BaystateHealth} as part of a QI initiative, we performed a retrospective study using Electronic Health Record information for adult inpatients from 2012-2016.  We sought to answer three questions:

\begin{enumerate}
\item Can we predict, directly from the EHR, which patients will satisfy a clinical definition of severe sepsis at various future times?

\item Can we use only unstructured text notes in the EHR for predictions?

\item How does prediction accuracy compare when using only unstructured data, only structured data, or both types of data from the EHR?

\end{enumerate}

\subsection{Previous Research}

Researchers use ``predict'' in  two different contexts.  Most are {screening} applications to determine whether patients \emph{currently} have sepsis, typically as judged by a gold standard team of physicians.   Others look at \emph{predicting ahead in time}, where sufficient structured data is not yet available to fulfill clinical criteria for severe sepsis.  Our focus here   is using {currently} available data to predict a {future} severe sepsis diagnosis, which is made using additional data available at that future time.

Bhattacharjee et al.\ [2017] describes  10 automated  tools using  structured data to predict patients with sepsis, severe sepsis, or septic shock.
None  make significant usage of free text notes, and most of these models are screening models.

A more recent paper by Horng et al.\ [2017] built Emergency Department triage screening models for sepsis (including severe sepsis and septic shock).  They make significant use of text notes and show about 25\%\ improvement when notes are added to structured data.  Text modeling involved term frequencies, bi-grams, and topic models, resulting in vectors with 15,000 dimensions.  (Reported results did not include  using unstructured data {alone}.)  

An earlier use of text notes to predict hospital mortality [Lehman 2012] employed Hierarchical Dirichlet Processes on UMLS clinical concepts extracted from the text.  Their mortality predictions showed text-based analysis was better than using structured data alone, and the combination of both produced best results (paralleling our results with severe sepsis).

In other work predicting sepsis, Desautels et al.\ [2016] reported on a machine learning approach using structured variables that showed improved predictive performance over a number of other methods for predicting sepsis, up to 4 hours in advance.  They use a recent definition of sepsis [Singer 2016] that differs from the traditional classifications of sepsis/severe~sepsis/septic~shock.
Paxton et al.\ [2013] built models to detect septic shock using 1011 structured data features.  They also discussed issues around using the Electronic Health Record for modeling in situations where some treatments have already started.

\section{Methods}

We analyzed 203,000 adult inpatient admissions (encounters) within {\Baystate} hospitals over 5 years from 2012 through 2016.  Severe sepsis patients were identified using a modified version of {\Baystate}'s clinical definition for severe sepsis, involving 8 structured variables.  Additionally, encounters were marked as severe sepsis if they had  corresponding ICD codes.  (Simply using severe sepsis ICD codes was not sufficiently reliable, as was previously found by Rhee et al.\ [2017].)  For each patient who satisfied this definition of severe sepsis, we used time stamps for the individual structured variables (or ICD codes) to compute the earliest time the severe sepsis definition was satisfied, which we term the \emph{Severe Sepsis Definition Time}.  

\subsection{Unstructured Data Models}

We collected textual  notes for patient encounters.  There were over 100 types of notes, with most being progress notes or history-and-physical notes.  When predicting severe sepsis 24 hours ahead for positive targets (severe sepsis patients), we removed all of the notes that occurred later than 24 hours prior to the Severe Sepsis Definition Time. Remaining data is referred to as the \emph{Modeling Data Window, MDW}. These encounters constituted positive examples for our modeling. 

For non-severe-sepsis patients, we chose a random time during their hospital stay and removed all notes that occurred later than 24 hours before that time.  These patients are negative examples.  Our intent was to simulate regular application of our models in a clinical setting, to suggest patients   for  further attention who are in danger of severe sepsis in the next 24 hours.

We eliminated encounters having no unstructured notes remaining in the MDW, because it makes no sense to try to predict with no information at all.  This left 68,482 total encounters, of which 1,427 (2.1\%)  satisfied the definition of severe sepsis during their stay.

We then concatenated all text information for an encounter\footnote{Not all types of text were used.} into a single text block which, along with the severe sepsis target flag, comprised our unstructured data.  Encounters were randomly divided into 3 groups of patients, stratifying them to maintain the global ratio of severe sepsis targets within each sample set.  For modeling and testing we used 3-fold cross validation, modeling on each set of 2 groups and using the remaining group to measure performance.  Final performance was computed over all 3 holdout sets.  We also built models using 2012--2015 data for model construction, and 2016 data for testing.

Models to predict 4 and 8 hours ahead used corresponding data preprocessing, with revised Modeling Data  Windows.  Note that models for 24-hour-ahead predictions had fewer encounters than 4 or 8 hour predictions, because the former had fewer non-empty Modeling Data Windows.  In particular, patients admitted with pre-existing severe sepsis were usually eliminated from modeling data for 24-hour models, because their diagnosis occurred in the first 24 hours of their encounter, leaving an empty MDW.

We represented unstructured data using 300-dimensional  GloVe vector embeddings for terms [Pennington 2014], and then summing the vectors for terms in the text.  Training was by ridge regression.

\subsection{Structured Data Models}

For models using only structured data, we selected 12 variables thought to be helpful for predicting Severe Sepsis and gathered their data, with timestamps, from the 
EHR. Values were computed for mean and standard deviation, plus counts of abnormal high, abnormal low, and normal readings, resulting in 29 modeling variables.  Once again, we discarded encounters that lacked data for all 29 variables, leaving no data in the Modeling Data Window.  

For combination models using both unstructured and structured data, we required that both types of data be non-empty in the MDW.

\section{Results}

Table~\ref{unstruct} gives summary results across all 3 cross-validation folds of the data using only unstructured EHR text.  To emphasize actionable results where there is a practical need to avoid over-alarming, we focused upon the most likely predicted 1\%, 5\%, and 10\% encounters for severe sepsis among hold-out data.

\begin{table}[htbp]
  \centering
  \caption{Predicting Severe Sepsis Using Only Text from the EHR}
  \label{unstruct}
    \begin{tabular}{rlrrrr}
    \multicolumn{1}{p{4.215em}}{\textbf{Predict}} &       & \multicolumn{1}{c}{\textbf{Encounters with}} & \multicolumn{1}{p{4.215em}}{\textbf{Top 1\% }} & \multicolumn{1}{p{4.215em}}{\textbf{Top 5\%  }} & \multicolumn{1}{p{4.215em}}{\textbf{Top 10\%  }} \\
	\multicolumn{1}{p{4.215em}}{\textbf{Ahead}} &       & \multicolumn{1}{c}{\textbf{Usable Data in}} &  \multicolumn{1}{p{4.215em}}{\textbf{Predicted }} & \multicolumn{1}{p{4.215em}}{\textbf{Predicted }} & \multicolumn{1}{p{4.215em}}{\textbf{Predicted }} \\
	 &       & \multicolumn{1}{c}{\textbf{Modeling Window}} & & & \\
          &       &       &       &       &  \\
    \multicolumn{1}{l}{\textbf{4 hours}} &       &       &       &       &  \\
          & Sample Size &                    129,421  &         1,294  &         6,471  &      12,942  \\
          & Targets found &                         2,527  & 521   & 801   & 952  \\
          & \% of Sample &       & \textbf{40\%} & \textbf{12\%} & \textbf{7\%} \\
          & \% of All Targets &       & \textbf{21\%} & \textbf{32\%} & \textbf{38\%} \\
          & AUC      & \textbf{0.636} &  &         &  \\
          &       &       &       &       &  \\
    \multicolumn{1}{l}{\textbf{8 hours}} &       &       &       &       &  \\
          & Sample Size &                    117,768  &         1,178  &         5,888  &      11,777  \\
          & Targets found &                         2,158  & 503   & 769   & 916  \\
          & \% of Sample &       & \textbf{43\%} & \textbf{13\%} & \textbf{8\%} \\
          & \% of All Targets &       & \textbf{23\%} & \textbf{36\%} & \textbf{42\%} \\
          & AUC   & \textbf{0.660} &       &       &  \\
          &       &       &       &       &  \\
    \multicolumn{1}{l}{\textbf{24 hours}} &       &       &       &       &  \\
          & Sample Size &                       68,482  &            685  &         3,424  &         6,848  \\
          & Targets found &                         1,427  & 412   & 707   & 829  \\
          & \% of Sample &       & \textbf{60\%} & \textbf{21\%} & \textbf{12\%} \\
          & \% of All Targets &       & \textbf{29\%} & \textbf{50\%} & \textbf{58\%} \\
          & AUC   & \textbf{0.727} &       &       &  \\
    \end{tabular}%
  \label{tab:addlabel}%
\end{table}%

Surprisingly, predicting ahead 24 hours gives better results than predicting ahead 4 or 8 hours.  This is a consequence of the former set of encounters being smaller in number, and having a higher percentage of longer-term patients who have more information available in their modeling windows.  This explanation was supported by building 4, 8, and 24~hour predictive models using the same 24-hour structured data, which showed improved performance for 4 and 8-hour predictions vs.\ 24-hour predictions.

We manually verified that unstructured text in the MDW did not already have sufficient information to \emph{diagnose} severe sepsis 24 hours prior to the Severe Sepsis Definition Time.  Looking at the top 1\% predicted encounters, we found very few (1\%-3\%) where narratives were sufficient for a prior diagnosis. We judged this to be an acceptable rate.  We similarly checked that Vasopressor drugs (an indication of septic shock) were not yet administered  during the Modeling Data Window.

Table~\ref{comparison} gives comparisons with unstructured, structured, or combination data when predicting 24 hours into the future.  Here we built severe sepsis models using encounters from 2012--2015, and testing with encounters from 2016.  Only encounters that have non-empty modeling windows for both unstructured and structured data were used.  The 2016 test set contained 13,603 usable encounters, for which 425 patients  (3.1\%) satisfied the severe sepsis criteria in the next 24 hours.  Using out-of-time test data further supported results from Table~\ref{unstruct} which used cross-validation.  To briefly summarize the comparison:  unstructured-only models performed comparably or slightly better than structured-only models, and the combination of unstructured and structured data gave a 5\%-10\% improvement over unstructured-only models.

\begin{table}[htbp]
  \centering
  \caption{Predicting Severe Sepsis 24 Hours in the Future Using Unstructured, Structured, or Combination EHR Data}
  \label{comparison}
    \begin{tabular}{p{10.0em}ccc|cc|cc}
    \textbf{Type of EHR Data} & \multicolumn{1}{p{2.0em}}{\textbf{AUC}} & \multicolumn{2}{c}{\textbf{Predicted Top 1\%}} & \multicolumn{2}{c}{\textbf{Predicted Top 5\%}} & \multicolumn{2}{c}{\textbf{Predicted Top 10\%}} \\
    \multicolumn{1}{c}{} &       & \multicolumn{1}{p{4.0em}}{{Number In Set}} & \multicolumn{1}{p{3.43em}|}{{Severe Sepsis}} & \multicolumn{1}{p{4.0em}}{{Number In Set}} & \multicolumn{1}{p{3.43em}|}{{Severe Sepsis}} & \multicolumn{1}{p{4.0em}}{{Number In Set}} & \multicolumn{1}{p{3.43em}}{{Severe Sepsis}} \\
    Unstructured Text only & 0.81  & 136   & 115   & 680   & 217   & 1,360 & 247 \\
    Structured Data only & 0.80  & 136   & 112   & 680   & 206   & 1,360 & 248 \\
    Both Unstructured Text   \\
	\multicolumn{1}{r}{And Structured Data} & 0.85  & 136   & 125   & 680   & 239   & 1,360 & 272 \\
    \end{tabular}%
  \label{tab:addlabel}%
\end{table}%

\section{Discussion}	

Our results indicate that we can construct models, directly from the Electronic Health Record, that are highly predictive  for severe sepsis diagnoses over the next 4, 8, and 24 hours,  and using only text notes.  This is the first result using text to \emph{predict ahead in time} a sepsis diagnosis, as well as the first sepsis predictions \emph{using only unstructured data}.

These severe sepsis models appear practically actionable and valuable.  For example when predicting ahead 24 hours, if we have 1,000 patients and take the model's top-scoring 10 patients (top 1\%), then we expect 6 of them to satisfy the severe sepsis definition in the next 24 hours.

It was somewhat surprising that such results were possible using merely our baseline ``bag of words'' distributed  representation for the text.  Future work will explore more advanced representations that explicitly represent negation [Gallant 2013], which is helpful in different modeling contexts (such as computer assisted coding of ICD-10 medical codes).

A possible source of noise in our data are the timestamps associated with both unstructured and structured items.  A planned prospective study will address this issue.

\subsection{The Case for Using Unstructured Text}

There are several advantages with using text notes for modeling, as compared with structured values.   The medical notes  tend to repeat lab results and structured variables, \emph{but only the important ones},  as judged by skilled clinicians.  This expert judgment constitutes a quite   valuable resource, and it appears only in the unstructured data.

Surprisingly, unstructured text requires less manual preprocessing effort than using structured data.  For example, with structured data there is not a {single} value for ``blood pressure'' to be extracted from the EHR; there are \emph{multiple} blood pressure readings, and these need to be rolled up using max, min, mean, deltas, etc.  Also, structured data presents more of a problem with missing data.

Another consideration is that we want to be able to predict many additional health-related targets, such as  re-admission risk, over sedation, CHF, etc.  Yet we cannot practically include \emph{all} structured data from the EHR --- a huge task --- so a separate subset of relevant structured variables needs to be manually defined and extracted for each different prediction target.  By contrast, we can  reuse \emph{the same set} of unstructured data for multiple prediction models.\footnote{An exception is that notes on patient history may be useful for targets like re-admission risk, but need to be excluded for  targets such as ICD-10  coding for current conditions.}

Sepsis may be an especially good candidate for using text notes because it has a complex definition involving many structured data values.  Structured data may work comparatively better for simpler targets that are defined by only a few structured data numbers.

\subsection{Concluding Remarks}

We will report additional details elsewhere, including results using a different prediction target (Over-Sedation), as well as results predicting severe sepsis using the MIMIC dataset [Johnson 2016].

Our findings strongly support the inclusion of unstructured notes when building predictive models.  Findings also support an implementation in a clinical setting to prospectively confirm accuracy and  usefulness of model predictions.

\subsubsection*{Acknowledgments}

Research supported by a grant from the National Science Foundation to MultiModel Research.  Baystate contributions were part of a QI initiative.  Thanks to Florin Brasoveanu, Patricia Humiston RN, Anees Hameed, Katie Devlin, Christian Lagier, Dan Greenberg and TechSpring for contributions to this work.

\section*{References}

\medskip

\small

Bhattacharjee P, Edelson DP, Churpek MM.  (2017) 
Identifying Patients With Sepsis on the Hospital Wards.
\emph{Chest}. 2017 Apr;151(4):898-907. 

Desautels, T., Calvert, J., Hoffman, J., Jay, M., Kerem, Y., Shieh, L.,  Shimabukuro D, Chettipally U, Feldman MD, Barton C, Wales DJ, Das, R. (2016). Prediction of Sepsis in the Intensive Care Unit With Minimal Electronic Health Record Data: A Machine Learning Approach. \emph{JMIR Medical Informatics}, 4(3), e28. http://doi.org/10.2196/medinform.5909.

Gallant, S. I. \&\ Okaywe, T. W. (2013)  Representing Objects, Relations, and Sequences.  \emph{Neural Computation} 25, 2038-2078.

Henry, K.E., Hager, D.N., Pronovost, P.J., \&\ Saria, S.  (2015) A targeted real-time early warning score (TREWScore) for septic shock. {\it Sci. Transl. Med.} 7, 299ra122.

Horng S, Sontag DA, Halpern Y, Jernite Y, Shapiro NI, Nathanson LA (2017) Creating an automated trigger for sepsis clinical decision support at emergency department triage using machine learning. PLoS ONE12(4): e0174708. https://doi.org/10.1371/journal.pone.0174708

Johnson AE, Pollard TJ, Shen L, Lehman LH, Feng M, Ghassemi M, Moody B, Szolovits P, Celi LA, Mark RG. (2016) MIMIC-III, a freely accessible critical care database. Sci Data. 2016;3:160035.

Ohno-Machado L (2011) Realizing the full potential of electronic health records: the role of natural language
processing. J Am Med Inform Assoc 18: 539. https://doi.org/10.1136/amiajnl-2011-000501
PMID: 21846784

Lehman LW, Saeed M, Long W, Lee J, Mark R (2012).  Risk stratification of ICU patients using topic models inferred from unstructured progress notes.  AMIA Annu Symp Proc.\ 2012;2012:505-11. Epub 2012 Nov 3.

Paxton, C., Niculescu-Mizil, A., \&\ Saria, S. (2013). Developing Predictive Models Using Electronic Medical Records: Challenges and Pitfalls. AMIA Annual Symposium Proceedings, 2013, 1109-1115.

Pennington, J., Socher, R. \&\ Manning, C. D. (2014). Glove: Global Vectors for Word Representation.. EMNLP (p./pp. 1532--1543).

Rhee, C.,  Dantes, R., Epstein, L., Murphy, DJ., Seymour, CW., Iwashyna, TJ., Kadri, SS., Angus, DC., Danner, RL., Fiore, AE., Jernigan, JA., Martin, GS., Septimus, E., Warren, DK., Karcz, A., Chan, C., Menchaca, JT., Wang, R., Gruber, S.\ and Klompas, M. (2017) "Incidence and Trends of Sepsis in US Hospitals Using Clinical vs Claims Data, 2009 - 2014," Journal of the American Medical Association, vol. 318, no. 13, pp. 1241-1249, 2017.

Singer M, Deutschman CS, Seymour CW, Shankar-Hari M, Annane D, Bauer M, Bellomo R, Bernard GR, Chiche J, Coopersmith CM, Hotchkiss RS, Levy MM, Marshall JC, Martin GS, Opal SM, Rubenfeld GD, van der Poll T, Vincent J, Angus DC.  (2016) The Third International Consensus Definitions for Sepsis and Septic Shock (Sepsis-3) \emph{JAMA}. 2016 Feb 23;315(8):801-10.

\end{document}